\documentstyle[prl,twocolumn,aps,tighten,floats,epsfig]{revtex}
\begin{document}
\draft 
\title{Wavefunction-based correlated {\em ab initio} calculations 
on crystalline solids} 
\author{Alok Shukla\cite{email}}
\address{Department of Physics, University of Arizona, Tucson, AZ 85721}  
\author{Michael Dolg, Peter Fulde} 
\address{Max-Planck-Institut f\"ur
Physik komplexer Systeme,     N\"othnitzer Stra{\ss}e 38 
D-01187 Dresden, Germany}
\author{Hermann Stoll} \address{Institut f\"ur Theoretische Chemie,
Universit\"at Stuttgart, D-70550 Stuttgart, Germany}

\maketitle

\begin{abstract}
We present a wavefunction-based approach to correlated {\em ab initio} 
calculations on crystalline insulators of infinite extent. 
It uses the representation of the occupied and the unoccupied (virtual) 
single-particle states of the infinite solid in terms of Wannier functions. 
Electron correlation effects are evaluated by considering virtual excitations
from a small region in and around the reference cell, keeping the electrons of
the rest of the infinite crystal frozen at the Hartree-Fock level. 
The method is applied to study the ground state properties
of the LiH crystal, and is shown to yield rapidly convergent results.
\end{abstract}
\pacs{ }

\section{INTRODUCTION}
\label{intro}
Density functional theory \cite{hkohn,skohn} (DFT) is a very powerful and
elegant approach to 
many-body 
systems. Its foundations are based upon the Hohenberg-Kohn (HK) 
theorem \cite{hkohn} which 
states that the total energy of the ground-state of a  many-particle
system 
in a given external potential can be expressed as a universal functional of 
its electronic charge density.
Thus, 
within a DFT-based formalism, one 
avoids constructing the many-body wavefunction of the system, and 
instead computes directly ground-state properties like the total energy,
magnetization, lattice constant etc.\ from its charge density. 
However, the exact form of the functional embodied in HK is unknown. 
Therefore approximations are required, the most important one being
the local-density approximation (LDA) developed by Kohn and Sham 
 \cite{skohn}. LDA has proven to be extremely successful 
for the ground-state properties of 
weakly correlated systems. 
However, for strongly correlated systems such as the ones containing f
electrons, its results are far from satisfactory. There have been many
improvements proposed to the LDA \cite{nlda},
but their applicability depends on the system involved, i.e., the present
DFT-based approaches are not amenable to {\em systematic} improvements.

The alternative approach to the problem of electron correlations 
aims at obtaining the many-particle wavefunction
of the system, by approximately solving the corresponding Schr\"odinger equation
within a given one-particle/many-particle basis set.
Since the Hamiltonian of the system under consideration is generally known
beforehand, there are no uncertainties of principle involved in this
approach. One can systematically improve the approach by enlarging the
basis set or by including more
terms in the expansion of the wavefunction of the system. However, such
"wavefunction-based" approaches scale quite unfavourably with the size of
the system. Therefore, in their {\em ab initio} form, they have been applied
mainly in quantum chemistry for small to medium-sized molecules. 
In condensed-matter physics, wavefunction-based many-body approaches 
are generally used in conjunction with model Hamiltonians,
particularly when the correlations are strong. Usually various approximations 
have to be made \cite{fulde} and an extension of {\em ab initio} quantum-chemistry 
type of methods to infinite systems did not seem feasible for a long time.

Two recent developments
may possibly bring {\em ab initio} wavefunction-based calculations for
infinite systems within the reach of routine computations: 
(a) advent of more powerful computers 
(b) use of local excitations in electronic structure calculations. 
The first of the aforesaid developments is technological in nature, and
its importance is self-explanatory. The second development is 
based on the fact that the correlation hole of an electron is a real-space
phenomenon. Local or localized orbitals can immediately tell us as to which
electrons will interact with each other strongly, and which ones
weakly. Although this has been realized long ago in solid-state
theory \cite{friedel} as well as quantum chemistry \cite{gilbert} only recently
have people started following it in full
earnest \cite{stol-ful,pulay,werner}. Realizing the potential behind the
localized-orbital-based approaches in the electronic structure theory of
solids, one of us developed the so-called ``incremental
scheme"  \cite{increment}.  
The incremental scheme is basically an expansion of the total corrrelation
energy per unit cell of a solid written in terms of interactions of
increasing complexity among the electrons assigned to localized 
orbitals (Wannier functions)
comprising the solid under consideration. 
Calculations based upon the incremental scheme have been performed
on a variety of solids \cite{inc-cal}.
However, there is one possible drawback associated with
the computations so far, namely, in all 
correlation calculations the infinite
crystal was modelled by a finite cluster. For the ionic systems the cluster
was embedded in a Madelung field, while for the covalent ones the saturation
of the  dangling bonds with hydrogen was the substitute for the environment.

The purpose of the present work is to study electron correlation 
effects in a crystalline insulator within the incremental scheme, {\em without}
compromising on the infinite nature of the system in any way. Thus by
comparing the results of the present work with calculations done by the
previous cluster-type approximation, we can critically examine both
approaches. Bulk LiH was chosen as a first test case for our method.
To the best of our knowledge this is the first calculation employing a 
wavefunction-based approach that does not truncate an infinite solid into
a finite cluster. Although the computational efforts involved is more than
for the cluster-based calculations, it is not excessive. Therefore, we believe,
that with further algorithmic improvements, the present approach can be
made viable for solids containing many atoms in a unit cell.

 The remainder of the paper is organized as follows. In section \ref{theory}
we briefly discuss the theoretical background of the present work.
In section \ref{results} we present the results of our calculations
performed on bulk LiH.
Finally, in section \ref{conclusion} we present our conclusions. 

\section{THEORY}
\label{theory}
The starting point of our approach is the representation of the
restricted Hartree-Fock (RHF) ground state of the solid in terms
of Wannier functions (WFs).
It is accomplished by solving the RHF equations of the infinite crystal
directly in the Wannier representation, as recently described  \cite{shukla}.
The applicability of the approach to periodic insulators of reduced dimensions
was demonstrated in another work of ours~\cite{shukla2}, where we
used it to study the ground states of an LiH chain and trans-polyacetylene,
at the RHF level. 
Assuming that the system consists of $N$ unit cells ($N \rightarrow \infty$)
each of which has $2n_c$ electrons, its RHF ground state will be described by 
$n_c$ doubly occupied WFs per unit cell. If we use 
Greek indices $\alpha$, $\beta$ etc. to denote the occupied
WFs localized in a given unit cell, the RHF wavefunction of 
a crystal 
is a Slater determinant composed of 
Wannier functions $\{ |\alpha({\bf R}_{j})\rangle; \alpha =1,n_{c}; j=1,N \}$.
The latter 
form an orthonormal set
\begin{equation}
\langle\alpha({\bf R}_i ) | \beta ({\bf R}_j) \rangle = \delta_{\alpha\beta} \delta_{ij}
\mbox{,}
\label{ortho}
\end{equation}
and Wannier functions localized in different unit cells are translated
copies of each other
\begin{equation}
|\alpha({\bf R}_{i}+{\bf R}_{j})\rangle = {\cal T} ({\bf R}_{i}) 
|\alpha({\bf R}_{j})\rangle \mbox{,}
\label{eq-trsym}
\end{equation}
where  the operator ${\cal T} ({\bf R}_{i})$ 
represents a translation by lattice vector ${\bf R}_{i}$. 
LiH has 2 occupied WFs per unit cell. Although, in our
approach, the WFs are assumed to be the linear combinations of Gaussian
lobe-type functions \cite{whitten}, the theory presented here is quite independent of
the choice of the basis set.

The next step is to choose a region of space --- henceforth to be 
called the ``correlation region''($\cal C$) --- in which the correlation
effects will be computed. The infinite region of space lying outside
the correlation region will be called the ``environment''($\cal E$). In the
correlation calculations, 
we consider virtual excitations from the occupied WFs of the correlation
region only, keeping the electrons of the environment ``frozen'' to the
occupied (RHF) space.
Because of the frozen character of the environment WFs, one can
immediately sum up their contribution to obtain an effective one-electron
potential called the ``environment potential''
\begin{eqnarray}
 U^{\mbox{env}}_{pq} & = & \sum_{\alpha({\bf R}_{j}) \in {\cal E} } 
( 2 \langle p \alpha({\bf R}_{j})|\frac{1}{r_{12}}|q \alpha({\bf R}_{j})
\rangle \nonumber \\
 &  & - \langle p \alpha({\bf R}_{j})|\frac{1}{r_{12}}|\alpha({\bf R}_{j}) q \rangle
 ) \; \mbox{,} \label{eq-uenv}
\end{eqnarray}
where $p$ and $q$ are two arbitrary basis functions and the factor of
two in the first term is due to the spin summation.
$ U^{\mbox{env}}_{pq}$
is evaluated using the Ewald summation technique \cite{inc-cal}, and one
is left with an effective Hamiltonian for the electrons located in the region $\cal C$. 
The physical interpretation of $U^{\mbox{env}}_{pq}$ is obvious --- it represents the 
influence of the electrons of $\cal E$ on those of $\cal C$.

The correlation contribution to the total energy per unit cell is
computed using the aforesaid incremental expansion \cite{increment}
\begin{equation}
\Delta E = \sum_i \epsilon_i + \frac{1}{2} \sum_{i \neq j} \Delta \epsilon_{ij}
            + \frac{1}{3!} \sum_{i \neq j \neq k} \Delta \epsilon_{ijk}
            + \ldots \; \mbox{,}
\label{incexp}
\end{equation}
where the summation over $i$ involves 
Wannier functions located in the reference cell, while those over $j$ and $k$
include all the Wannier functions of the crystal. The $\epsilon_i$ 
(``one-body'' increment)
are computed by considering virtual excitations only from 
the $i$-th Wannier function, freezing the rest of the solid at the
HF level. The ``two-body'' increment $\Delta \epsilon_{ij}$ is defined as
$\Delta \epsilon_{ij}= \epsilon_{ij} - \epsilon_i - \epsilon_j$
where $\epsilon_{ij}$ is the correlation energy of the system obtained
by correlating two distinct Wannier functions $i$ and $j$. Thus  
$\Delta \epsilon_{ij}$ represents the correlation contribution 
of electrons localized on two ``bodies'' $i$ and $j$. Similarly 
higher-order increments are defined.
For ionic systems such as LiH, generally one- and two-body increments account
for the bulk of correlation contribution --- three-body increments being
negligible. 
The incremental expansion of Eq.(\ref{incexp}) is completely independent of 
the method of computation.
Since LiH is a small system, for the present calculations of increments it was
possible 
to use the full configuration-interaction (CI) approach which in the physics
literature is known as the exact diagonalization method. For larger systems,
however, full-CI will not be feasible. In that case one can resort to any of
the other available size-extensive correlation schemes \cite{fulde}.

We finally comment upon the construction of the space of virtual orbitals
used to compute the correlation effects. We find it absolutely essential that 
the virtual orbitals be also localized. For this purpose we adopted the 
approach suggested by Pulay \cite{pulay}. It consists of first 
orthogonalizing the basis functions localized in the reference cell --- using
projection operators --- to the occupied Wannier functions of the reference cell,
and to those located in up to third-nearest neighbor unit cells. This yields a
set of virtual functions orthogonal to the occupied space, but not orthogonal
amongst each other. Orthogonality in the virtual space 
is finally achieved by a symmetric orthogonalization procedure which does not
destroy the local character of the functions. In the end, we obtain virtual 
Wannier functions which are somewhat delocalized due to their orthogonalization
tails, however, quantitatively the delocalization is not very significant.
Since occupied and virtual orbitals in our approach form an orthonormal
set, almost the whole machinery of electron correlation treatments developed during
the last decades in {\em ab initio} quantum chemistry for atoms and molecules can 
also be applied to treat polymers or solids.

\section{Calculations and Results}
\label{results}
Among the recent studies of bulk LiH, the most noteworthy are HF level studies
of Dovesi et al. \cite{hf-dovesi} and of Bella\"{\i}che and L\'{e}vy \cite{hf-bel},
and DFT-based studies of Bella\"{\i}che and coworkers \cite{dft-kunc,dft-bel}.
For the sake of comparison, in our study we used the Gaussian lobe representation of 
the polarized contracted basis sets used by Dovesi et al. \cite{hf-dovesi}, i.e.,
[3s,1p] on hydrogen and [2s,1p] on lithium.
The experimental fcc geometry of the system was assumed with hydrogen
on the $(0,0,0)$ position and Li on $(0,0,a/2)$ position, $a$ being the lattice
constant. First all-electron Wannier HF calculations were performed at fifteen
lattice constants in the range 4.0 --- 4.225 \AA. The results provided us
with the optimized lattice constant and cohesive energy of LiH at the HF level.
From the orbitals of these calculations the appropriate environment potentials 
were constructed (cf. Eq.(\ref{eq-uenv})), which were subsequently
used to perform incremental-scheme-based correlated calculations, 
employing the full-CI code of Knowles and Handy \cite{knowles}.
At the correlated level we resorted to the frozen-core approximation by
including the Li Wannier functions of region $\cal C$ also in
$U^{\mbox{env}}$. Since Li 1s type orbital is highly localized and core-like,
this approximation is quite reasonable and is analogous in spirit to the
pseudopotential calculations performed routinely in condensed-matter physics.

In table \ref{tab-comp} we present the results of incremental calculations
performed for the infinite solid for the lattice constant of 4.07 \AA. 
The increments reported there include the one-body increment corresponding 
to the correlation energy of the polarized hydrogen anion of the reference unit
cell and the two-body increments involving the interaction of the reference
hydrogen anion with those situated in up to third-nearest neighbor unit cells. 
Note the rapid decrease of the two-body increments with increasing distance between the ions 
which is of van der Waals type. The reason behind this can be intuitively
understood as follows. When, for the purpose of the computation of correlation
effects, atoms (or ions) are virtually excited, they develop local (virtual)
electric dipole moments. Therefore, the two-body increment corresponds to the 
interaction of two localized (virtual) dipole moments, the interaction
between which is nothing but the van der Waals interaction. A noteworthy point
is that localized orbitals are absolutely essential for this argument
to hold and, within a Bloch orbital approach, such a simplification is
not possible. For the sake of comparison, in table \ref{tab-comp} we also
present the results of calculations performed on a finite cluster modelling
the bulk LiH, using the same basis set and geometry. The finite
cluster calculations were performed with MOLPRO program \cite{molpro}
and include only the most important increments namely one-body and the
nearest-neighbor two-body increments. The maximum disagreement of 
$\approx 1.6\times10^{-3}$ Hartrees between the two calculations is very encouraging, 
and essentially defines the error bars of the numerics.

The equilibrium lattice constant, cohesive energy and the bulk modulus
obtained from our calculations --- both at the HF and correlated level --- are 
presented in table \ref{tab-opt} together with results of other 
authors \cite{lih-exp-a,lih-exp-ce,lih-exp-b,lih-exp-c,lih-exp-d,gerlich,roma}.
Before accounting for the zero-point
vibrations of the nuclei, the DFT-based calculations systematically
underestimate the lattice constant. Since the hydrogen nucleus is very light,
one would expect the energy of its zero-point vibrations to make non-negligible
contributions to the total energy. Indeed, Roma et al. \cite{roma} applying
density functional perturbation theory (DFPT) found that the inclusion of
the zero-point energy increases the equilibrium lattice constant of LiH by 0.08 \AA.
Bella\"{\i}che et al. \cite{dft-bel}, by adding the lattice-constant-dependent
zero-point energies obtained by Roma et al. \cite{roma}, to their previously derived DFT energies \cite{dft-kunc}
were able to obtain near-perfect agreement with the experimental lattice constant
in case of the Wigner expression for the correlation energy, whereas their result
is still 0.06 \AA \ too small for the more modern Perdew-Zunger parametrization
of the Ceperly-Alder quantum Monte Carlo correlation energy.
For the wavefunction-based approaches table \ref{tab-opt} shows that the HF
approximation yields a too large lattice constant due to an overestimation of
the ionic radii, particularly for the anion.
Once electron correlation effects are taken into account, we get an agreement of
better than 0.01 \AA \ with the experimental value of the lattice constant,
without accounting for zero-point effects.

The experimentally measured cohesive energy of LiH, corresponding to its
dissociation into Li$^+$ and H$^-$, is 0.346 a.u. \cite{lih-exp-ce}.
This, when corrected by the difference of the electron affinity of H and the
ionization potential of Li, leads to a cohesive energy of 0.176
a.u. with respect to the separation into free Li and H atoms.
In the present case, at the correlated level, the  Li core
is held frozen. Thus the theoretical cohesive energies both at the HF and the
correlated level are obtained by subtracting from the total energy per unit cell,
the HF limits of the energies of free Li (7.4327 a.u. \cite{clementi})
and H (0.5 a.u.) atoms. From table \ref{tab-opt} it is obvious that
for the HF cohesive energy there is very good agreement among the results
of all authors. The difference of 0.0006 a.u. between our results and
the one of Dovesi et al. \cite{hf-dovesi}
is possibly due to our use of Gaussian lobe functions instead of Cartesian Gaussians.
By subtracting our HF cohesive energy (0.1294 a.u.) from the experimental one (0.1760 a.u.),
one gets $\approx 0.0466$ a.u. as the ``experimental'' value of the correlation contribution
to the cohesive energy per unit cell. The correlated cohesive energy that we obtain in the present
work is 0.1644 a.u. and corresponds to 93.4 \% of the experimental value.
The calculated corresponding correlation contribution of 0.0350 a.u. is
75.1 \% of the experimental value. 

In order to account for a higher percentage of the
correlation contribution we would need to apply larger basis sets including additional diffuse
and also higher angular momentum functions. Unfortunately, since our program is still under
development, such large-scale calculations are not possible at present.
Nevertheless, one can estimate the effect of the basis sets incompleteness
for the time being from finite cluster calculations using extended basis sets \cite{basis}.
We obtained a value of -0.035486 a.u. for the one-body contribution and -0.001544 a.u. 
for the nearest-neighbor two-body contributions. Since we are interested in the values of the 
corresponding contributions obtained with the extended basis sets for the infinite solid 
within the Wannier function approach, we scale these numbers according to the ratios of the 
corresponding numbers in the two columns of table \ref{tab-comp} and obtain
the estimates -0.033652 a.u. and -0.001230 a.u., respectively.
By comparing these values to the corresponding numbers obtained with the smaller basis set 
for the bulk case, we obtain  corrections of -0.004335 a.u. and -0.000553 a.u. for the 
one-body and two-body term, respectively, leading to a total correlation energy gain of 
-0.0110 a.u. Our so-corrected estimate of the correlation contribution to the 
cohesive energy per unit cell of 0.0460 a.u. (0.0350 a.u. + 0.0110 a.u.) is
in almost perfect agreement with the corresponding experimental number 
($\approx$ 0.0466 a.u.). We want to point
out here, that further technical improvements in our implementation of the method
will make calculations with extended basis sets for the infinite system feasible
and therefore extrapolations from finite cluster calculations unnecessary.
The fact that an extension of the basis sets leads to a calculated value
essentially in agreement with the experimental result underlines the potential
of the proposed method, i.e., the possibility of systematic improvement.

The results of our Wannier-function-based calculations, using the smaller basis
set of Dovesi et al. \cite{hf-dovesi}, show that in a sufficiently wide range of values 
around the equilibrium lattice constant (15 points in the interval 4.0 \AA \ $\le$ $a$ $\le$ 4.225 \AA) 
to high accuracy (correlation coefficient 0.999983) the correlation energy depends {\em linearly} on
the lattice constant $a$ ($E_{corr}(a.u.) = -0.034030 + 0.002935 *(a (\AA) - 4.06)$).
This is in agreement with an assumption made in previous work on NiO by Doll et al. \cite{inc-cal}.
Such a ``linear law'' is important because it allows one to perform correlated calculations
only for two values of the lattice constant near its equilibrium value and to
deduce the other values by interpolation. Assuming also a linear behavior of
the correlation contributions obtained with the extended basis sets mentioned above 
\cite{basis}, we derive an estimated equilibrium lattice constant of 4.054 \AA \ compared
to the actually calculated value of 4.067 \AA \ and the experimental result of 4.060 \AA.

Finally, we discuss the influence of correlations on the bulk modulus of LiH.
The experimentally measured values reported in literature \cite{lih-exp-b,lih-exp-c,lih-exp-d}
are actually obtained by fitting a pressure-volume (p(V)) relationship to experimental
data points. The values for $^7$LiD are only about 1 \% to 3 \% larger than for $^7$LiH \cite{lih-exp-b,roma}
and sometimes a common value is reported \cite{lih-exp-d}.
Depending on the chosen ansatz for the p(V) equation of state (Birch, Murnaghan, Vinet)
the values reported for 300 K range from 31.1 GPa to 36.2 GPa, with error bars of 1 \% to 6 \%. 
In addition to this uncertainty, the values have to be corrected to 0 K, which roughly corresponds 
to an increase by 6 \% \cite{lih-exp-c,gerlich}, i.e., a range from 33.0 GPa to 38.4 GPa.
A direct comparison of our calculated values, which do not rely on any fitting of
an equation of state and are for 0 K, and the effective experimental parameters is to be viewed 
with care. From table \ref{tab-opt} it is obvious that our HF value of 33.4 GPa is in
within the interval of experimental values, whereas our correlated (estimated)
 value of 
36.5 (37.9) GPa appears to be slightly too large.
Roma et al. \cite{roma} observed a reduction of the bulk modulus by 4 GPa ($\approx$ 10 \%) due 
to zero-point effects. Using their zero-point corrections Bella\"{\i}che et al. \cite{dft-bel} 
were able to obtain close agreement with the experimental values of the bulk modulus in case of 
the Wigner correlation energy expression (33.9 GPa), whereas for the Perdew-Zunger parametrization 
also a relatively large value was obtained (36.1 GPa). We note that by adding the
corrections calculated by Roma et al. \cite{roma} to our energy-volume data, we end up with correlated
results which are not in good agreement with the experimental values, i.e., a lattice constant of
4.143 \AA \ and a bulk modulus of 30.7 GPa. Since our uncorrected
values appear to be stable with respect to systematic improvements of the calculation and are
also quite close to the experimental values, we believe that the zero-point corrections obtained
by Roma et al. may be slightly too large, despite the fact that previous estimates were even 
significantly larger \cite{martins}.

\section{Conclusions and Future Directions}
\label{conclusion}
In conclusion, a new wavefunction-based approach has been presented, which
allows the {\em ab initio} determination of electron correlation effects
in a crystalline insulator, without abandoning its infinite character. 
The approach merges ideas from quantum chemistry and solid-state physics,
and has the benefit of being amenable to systematic improvements of results.
It has been shown to be viable by a first successful application to an ionic
crystal, namely bulk LiH. The applicability of the present approach
to covalent insulators and semiconductors such as diamond, Si, GaAs etc. 
is essentially straightforward. The only trivial modification in those
cases will be that the occupied (and virtual) single-particle states will 
consist of bond-centered Wannier functions, instead of the atom-centered
Wannier functions encountered in case of ionic systems.
As a matter of fact, the viability of the present local approach to electron
correlations in covalent systems has already been demonstrated by
Paulus et al.~\cite{inc-cal}---albeit within a finite-cluster 
approximation---in their study of Group IV semiconductors. Its generalization
within the infinite-crystal method presented in this work will be the
subject of a future investigation. 

As far as the treatment of excited states of crystals is concerned, the
quasi-particle excited states such as an electron (hole) in the conduction
(valence) band can be treated straightforwardly using the present approach,
as these excited states can be regarded as the ground states of $N+1$ ($N-1$)
electrons, where $N$ is the total number of electrons in the unperturbed
system. Of course, the electron correlation effects in these cases will be
more complicated as compared to the $N$-particle ground state, owing to 
the polarization cloud of the extra electron (hole) in the system.
Applicability of this approach to quasi-particle excitations was demonstrated
by Gr\"afenstein et al.~\cite{graef1,graef2}, who computed the
correlated dispersion relations of holes in the valence bands of
Group IV semiconductors. Its generalization to the infinite crystal, as
well as extension to the conduction-band dispersion relations, will
also be investigated in future.

As far as the general computational viability of our approach is
concerned, undoubtedly, at present, it is more demanding than the DFT-based 
approaches. However,  we believe that by utilizing modern order-$N$ type
methods of electronic-structure theory, it can be made competitive with 
such methods. Conceptually speaking, there is nothing in our
approach that restricts its applicability to systems containing unit cells
of certain shapes, sizes, or number of electrons. The frontiers in these areas
 will be decided
by the progress in the computer hardware, and by the improvements in the 
algorithms of electronic structure theory. However, in order to be more
specific about the shapes and sizes of the systems which can be studied using 
the present approach, we believe that more experience with it is essential.
Work along these directions is presently underway in our group.
\acknowledgements
The authors are grateful to Dr. G. Roma for providing the actual data for 
the zero-point energy corrections plotted in Ref. \cite{roma}.

\clearpage
\newpage
%
%
\begin{table}  
 \protect\caption{Various increments to the correlation energy in Hartrees of
 bulk LiH computed by the Wannier-function-based approach presented in this
 work.  For comparison, the results obtained using the method of Doll et
 al.  \protect\cite{inc-cal} are presented, for the most important increments,
 under the heading of Finite Cluster. NN stands for nearest neighbors. All the
 calculations were performed at the lattice constant of $4.07$ \AA.} 
\protect\begin{center}  
  \begin{tabular}{lll} \hline 
 \multicolumn{1}{c}{Correlation Increment}   &
 \multicolumn{1}{c}{Wannier Function} 
   & \multicolumn{1}{c}{Finite Cluster} \\ \hline
 one-body       & -0.029317 & -0.030915 \\
 two-body (1NN) & -0.000677 & -0.000850 \\
 two-body (2NN) & -0.000113 &  ---   \\
 two-body (3NN) & -0.000023 &  ---  \\ \hline
   \end{tabular}                      
   \end{center}  
  \label{tab-comp}    
\end{table}  
\begin{table}  
 \protect\caption{Equilibrium values of various quantities for the bulk
LiH obtained in this work as compared to those of other authors, and the
experiment.  Quantities under the column DFT refer to slightly different
exchange-correlation functionals used (see Ref.\protect\cite{dft-bel} for details).
The theoretical results presented below do not include the effect
of zero-point vibrations, unless specified otherwise. Lattice constants are in 
\AA, energies in a.u. (i.e., Hartree) and the bulk moduli are in GPa.}
\protect\begin{center}  
  \begin{tabular}{lllllll} \hline 
 \multicolumn{1}{c}{Quantity}   
 & \multicolumn{1}{c}{HF$^a$} 
 & \multicolumn{1}{c}{HF$^b$} 
   & \multicolumn{1}{c}{HF$^c$}
   & \multicolumn{1}{c}{Correlated$^a$}
   & \multicolumn{1}{c}{DFT$^d$}
   & \multicolumn{1}{c}{Experiment}
 \\ \hline
 Lattice Constant & 4.106 & 4.102 & 4.110 & 4.067 & 3.91--3.98     & 4.060$^f$ \\
                  &       &       &       &       & 3.99--4.05$^e$ & \\
 Cohesive Energy  &0.1294 & 0.1302& 0.1305& 0.1644&  ---           & 0.1760$^g$ \\
 Bulk Modulus     &  33.4 & 34.1  & 35.0  & 36.5  & 33.9--36.1$^e$ & 33.0--38.4$^h$ \\
                  &       &       &       &       &                & 31.3--36.2$^i$ \\
  \hline
   \end{tabular}                      
   \end{center}  
$^a$ This work \\
$^b$ Ref.\protect\cite{hf-dovesi} \\
$^c$ Ref.\protect \cite{hf-bel} \\
$^d$ Ref.\protect \cite{dft-bel}, only all-electron results are considered. \\
$^e$ Includes the effect of zero-point motion. \\
$^f$ Extrapolated T=0 value \cite{lih-exp-a} \\
$^g$ The cohesive energy reported here is 
the ``atomic'' cohesive energy obtained by correcting the ``ionic''
cohesive energy reported in ref. \cite{lih-exp-ce}. See text for
an explanation. \\
$^h$ estimated for 0 K by scaling 300 K values with 1.06 \cite{lih-exp-c,gerlich} \\
$^i$ 300K \cite{lih-exp-b,lih-exp-c,lih-exp-d}.
  \label{tab-opt}    
\end{table}  

\begin{references}
\bibitem{basis}{The basis set for hydrogen was [6s,5p,4d] and the
one for Li was [5s,4p] type. These basis sets are, respectively,
referred to as AV5Z and VQZ and were taken from the MOLPRO program
library \cite{molpro}.}
\end{references}
\end{document}